\documentclass[12pt]{amsart}
\usepackage{amsfonts, amsmath, amssymb}
\usepackage{amsthm}
\usepackage{graphicx}
\usepackage{float}
\usepackage{enumitem}

\usepackage[utf8]{inputenc} 
\usepackage{enumitem}
\usepackage{lipsum}  

\usepackage{graphicx}
\usepackage{subcaption}
\usepackage[utf8]{inputenc}
\usepackage{url}

\newtheorem {Theorem}{Theorem}
\numberwithin{Theorem}{section}

\theoremstyle{definition}

\theoremstyle{remark}
\newtheorem{Remark}[Theorem]{Remark}

\numberwithin{equation}{section}
\expandafter\chardef\csname pre amssym.def
at\endcsname=\the\catcode`\@ \catcode`\@=11
\def\undefine#1{\let#1\undefined}
\def\newsymbol#1#2#3#4#5{\let\next@\relax
 \ifnum#2=\@ne\let\next@\msafam@\else
 \ifnum#2=\tw@\let\next@\msbfam@\fi\fi
 \mathchardef#1="#3\next@#4#5}
\def\mathhexbox@#1#2#3{\relax
 \ifmmode\mathpalette{}{\m@th\mathchar"#1#2#3}%
 \else\leavevmode\hbox{$\m@th\mathchar"#1#2#3$}\fi}
\def\hexnumber@#1{\ifcase#1 0\or 1\or 2\or 3\or 4\or 5\or 6\or 7\or 8\or
 9\or A\or B\or C\or D\or E\or F\fi}

\font\teneufm=eufm10 \font\seveneufm=eufm7 \font\fiveeufm=eufm5
\newfam\eufmfam
\textfont\eufmfam=\teneufm \scriptfont\eufmfam=\seveneufm
\scriptscriptfont\eufmfam=\fiveeufm

\catcode`\@=\csname pre amssym.def at\endcsname

\begin{document}





\title[]{Assessing the efficiency of different control strategies for the coronavirus (COVID-19)  epidemic }

\author[C. Castilho et al.]{C\'esar Castilho$^1$, João A. M. Gondim$^2$, Marcelo Marchesin$^3$, Mehran Sabeti$^4$}

\email{castilho@dmat.ufpe.br}

\bigskip

\maketitle

\centerline {$^1 \, $Departamento de Matem\'atica} \par \centerline{
Universidade Federal de Pernambuco} \par  \centerline{Recife, PE
CEP 50740-540 Brazil} \vspace{0.2cm}

\centerline {$^2 \, $Unidade Acadêmica do Cabo de Santo Agostinho} \par \centerline{
Universidade Federal Rural de Pernambuco} \par  \centerline{Cabo de Santo Agostinho, PE
CEP 54518-430 Brazil} \vspace{0.2cm}

\centerline{$^3 \,$Departamento de Matem\'atica}
\par \centerline{Universidade Federal de Minas Gerais} \par  \centerline{Belo Horizonte, MG CEP 31270-000 Brazil} \vspace{0.2cm}

\centerline {$^4 \,$Instituto de Ciências Exatas e Tecnológicas} 
\par \centerline{Universidade Federal de Vi\c cosa Campus-Florestal} \par  \centerline{Vi\c cosa, MG
CEP 35690-000 Brazil} \vspace{0.2cm}

\begin{abstract}
The goal of this work is to analyse the effects of control policies for the coronavirus (COVID-19) epidemic in Brazil. This is done by considering an age-structured SEIR model with a quarantine class and two types of controls. The first one studies the sensitivity with regard to the parameters of the basic reproductive number $R_0$ which is calculated by a next generation method. The second one evaluates different quarantine strategies by comparing their relative total number of deaths.
\end{abstract}

\vspace{0.4cm}
\centerline{ {\bf Key Words:} 
Coronavirus, Quarantine, Epidemic, SEIR models} \vspace{0.4cm} 

\newpage 

\section{Introduction}
At the ending of  $2019$ several cases of pneumonia of unknown
etiology were  detected in Wuhan City in the Chinese province of
Hubei. The Chinese Country Office of the World Health Organization
\footnote{World Health Organization Website \url{https://www.who.int/}}was
informed and reported that a novel coronavirus (officially named
COVID-19) was identified on January $7^{th}$ as the cause of such
infection. The imminent potential for worldwide spread was soon
recognized and an international alert was issued.\par 
 COVID-19 was shown to be very lethal and easily spreading. China's effort to mitigate the harm were apparently quickly taken yet as many as more than $75,000$
infected cases were reached in Wuhan before the end of January \cite{wu2020nowcasting}. Due to the highly interconnected world
we presently live in, the disease quickly spread outside China
reaching practically every country in the world with several
different degrees of seriousness. On March $11^{th}$, due to the
seriousness of the situation, the WHO declared it a Pandemic. \par

The goal  of this study is to assess through the analysis of a
differential equations  model the importance  of different 
control policies for the Brazilian COVID-19 epidemic. Even though Brazil is considered for the scope of this paper, the techniques and tools used in this study can be easily adapted for any other country. The impact of different control strategies are
qualitatively evaluated and mathematically based guidelines concerning different  
protective measures and quarantine strategies are formulated.  The paper is organized as follows: In Section \ref{section:The Age-structured SEIR model}, the age-structured SEIR model with quarantine is formulated. Demographic data from Brazil is introduced and discussed. 
In Section \ref{section:The unstructured SEIR model}, the classical SEIR model without vital dynamics and with a quarantine compartment is studied. The goals here are, firstly, to adjust parameters and to fit the real data, and secondly, to study the necessary quarantine efforts and times so to be able to influence the epidemic. In Section \ref{section:Control strategies for the age-structured model}, the parameters for the age structured model are adjusted (using the ones calculated on the previous section). The next generation approach is used to calculate the basic reproduction number and a sensibility analysis is carried on. In Section \ref{section:The effects of different quarantine policies}, different quarantine strategies are considered and compared. We draw our conclusions in \ref{ref:conclusions}.

 \section{The Age-structured SEIR model}
 \label{section:The Age-structured SEIR model}
 A classical SEIR model is used with the addition of a quarantine class as proposed in \cite{jia2020modeling}. Since age is an important factor on the COVID-19 epidemic, it will be assumed that the population is age structured (see \cite{inaba2006mathematical}, \cite{thieme2001disease}, \cite{castillo1989epidemiological} for continuous models and \cite{zhou2019global}, \cite{zhou2004dynamics} for discrete models).   Three age classes are used; $i=1 :$ infants with ages in the interval $[0,19]$,   $i=2 :$  adults with ages in the interval $[20,59]$, and $i=3 :$ elderly with ages in the interval $[60,100]$. The proportion of each age class in the Brazilian population is shown in Table \ref{t-age_classes} (see \cite{ibge}).  \par 
 
 Let   $S_i(t)$, $E_i(t)$, $I_i(t)$, $R_i(t)$ and $Q_i(t)$ represent the number of susceptibles, exposed, infected, removed and quarantined at age class $i$ respectively at time $t \ge 0$. 
  The equations are as follows
 \begin{equation}
 \begin{aligned}
 Q^{\, \prime}_i(t) &= p_i \, S_i(t) - \lambda_i \, Q_i(t) \, , \, \, \, i=1,2,3. \, \, \, , 
 \\[0.4cm]
 S^{\, \prime}_1(t) &= \Lambda - (\mu_1 + \rho_1) \, S_1(t)  - S_1(t) \, \left( \sum_{j=1}^3 \beta_{1j} \, S_j(t)\right)\\  &- p_1 \, S_1(t) + \lambda_1 \, Q_1(t) \, ,\\[0.4cm]
 S^{\, \prime}_2(t) &= \rho_1 \, S_1(t) - (\mu_2 + \rho_2) \, S_2(t)  - S_2(t) \, \left( \sum_{j=1}^3 \beta_{2j} \, S_j(t) \right)\\  &- p_2 \, S_2(t) + \lambda_2 \, Q_2(t) \, ,\\[0.4cm]
 S^{\, \prime}_3(t) &=\rho_2 \, S_2(t) - \mu_3 \, S_3(t)  - S_3(t) \, \left( \sum_{j=1}^3 \beta_{3j} \, S_j(t)\right)  \\ &- p_3 \, S_3(t) + \lambda_3 \, Q_3(t) \, ,\\[0.4cm]
 E^{\, \prime}_i(t) &=   S_i(t) \, \left( \sum_{j=i}^3 \beta{ij} \, I_j(t) \right) - (\sigma_i + \mu_i) \, E_i(t) \, ,\\[0.4cm]
 I^{\, \prime}_i(t) &=  \sigma_i \, E_i(t)  - (\gamma_i + \mu_i + m_i ) \, I_i(t) \, \, ,\\[0.4cm]
 R^{\, \prime}_i(t) &= \gamma \, I_i(t) - \mu_i \, R_i(t) \, , \\
 \end{aligned}
 \label{eq:seir_estruturado}
 \end{equation}
 The parameters are all non-negative (or positive) and are described in Table~\ref{t-parametrosSIRBasico}.  $p_i$ and $\lambda_i$ are the quarantine entrance and exit rates for class $i$, respectively. $\Lambda \, , \, \mu_i $ and   $\rho_i$ are the vital parameters. In the disease free situation the population is assumed to be at demographic equilibrium. $\gamma_i$ is the recovery rate, $m_i$ the disease induced death rate and $\beta_{ij}$ is the infection rate between class $i$ and class $j$. Typically, it will be assumed that $\beta_{ij}=\beta_{ji}$. \par 
 The class $Q_i$ has the effect of removing susceptible individuals from the infection dynamics. If
 $p_i=\lambda_1=0$ there is no quarantine and the system reduces to an age structured SEIR model. 
 \begin{table}
 	\centering
 	\caption{Age Classes.}
 	\label{t-age_classes}
 	\begin{tabular}{ccccc}
 		\hline  
 		\\
 		Class 	  &  Age (years) & $\%$ Population & $\%$ Mortality (year)  & $\mu_i $ (year)  \\ 
 		\\ \hline
 		\\
 		1 & [0,19] & 40.2 $\%$  & 12.6 $\%$ & $1.959 / 1000 $	\\
 		\\
 		2 & [20,59] & 50.5 $\%$ & 33.2 $\%$	& $4.109  / 1000 $\\
 		\\
 		3 & [60,100] & 9.3 $\%$ & 54.2 $\%$ & $36.425 / 1000 $ \\
 		\\ \hline
 	\end{tabular}%
 \end{table}

 According to \cite{ibge} Brazil has  18,67 births and 6.26 deaths by 1000 inhabitants per year, giving an annual population growth of 1.24 $\%$. Let $N$ denote the total population and $D$ the total deaths per year, thus
 $$ \frac{D}{N} = \mu = \frac{6.25}{1000} \, .$$ 
 Similarly, let $D_i$ and $N_i$ be the  number of deaths per year and $N_i$ be the population of age class $i$ respectively. Thus
 $$  \mu_i =\frac{D_i}{N_i}\, . $$
 With this notation the data on Table \ref{t-age_classes} is denoted by
 $$ \% \, \mbox{Population}= \frac{N_i}{N} \, \, \, \mbox{and} \, \, \,  \% \, \mbox{Mortality}= \frac{D_i}{D} \, . $$
 $\mu_i$ is calculated by
 $$ \mu_i = \frac{D_i}{N_i} =  \frac{D_i}{D} \, \frac{D}{N} \, \frac{N}{N_i} =  \mu \, \frac{D_i}{D} \, \frac{N}{N_i} = \mu \, \frac{\left( D_i/D \right) }{\left( N_i/N \right) } \, . $$

  \par
  The disease free steady state is denoted by
  \begin{equation} 
  \label{eq:free_disease}
   S_1^* \, , \, S_2^* \, , \, S_3^* \, , \, E_i=I_i=R_i=0 \, \, \, i=1,2,3 \, \end{equation}
  where by $S_i^*$ we denote the number of individuals of age class $i$ (all individuals are susceptibles, see table \ref{t-age_classes}).
  For the model without quarantine, adding the equations for the disease free state gives
  $$ (S_1(t) + S_2(t) + S_3(t))^{\, \prime}(t) = (\Lambda - \mu_1 \,S_1(t) - \mu_2 \, S_2(t) - \mu_3 \, S_3(t)) \, . $$
   Assuming that the total population  is constant and on demographic equilibrium, using the values for the population distribution as the equilibrium values, one must have 
  $$\Lambda = \mu_1 \, S_1^* + \mu_2 \, S_2^* + \mu_3 \, S_3^* = 6.25/1000 \, . $$
  The actual annual growth rate will be ignored. Since the time frame of interest is small compared to the demographic time scale, this has no consequences on the main conclusions of this work. The demographic equilibrium implies that $\rho_1$ and $\rho_2$ satisfy
  $$ \rho_2= \frac{\mu_3 \, S_3^*}{S_2^*} = 6.707 \times 10^{-3} \, \, \, \mbox{and} \, \, \, \rho_1 = \frac{(\mu_2 + \rho_2 ) \, S_2^*}{S_1^*} = 11.033 \times 10^{-3} \, .$$
  
 \begin{table}
 	\centering
 	\caption{Parameters of the basic SEIR model with vital dynamics.}
 	\label{t-parametrosSIRBasico}
 	\begin{tabular}{cl}
 		\hline  
 		\\
 		Parameter 	  &  Description \\ 
 		\\ \hline
 		\\
 		$p_i$ & quarantine entrance rate for class $i$.	\\
 		\\
 		$\lambda_i$ & quarantine exit rate for class $i$.	\\
 		\\
 		$\Lambda$ & recruitment rate.	\\
 		\\
 		$\mu_i$ & natural death rate for class  $i$.	\\
 		\\
 		$\rho_i$ & survival rate for class $i$	to class $i+1 \, \, \, i \le 2$. \,\\
 		\\
 		$\beta_{ij}$ 	    &  pathogen's transmission rate between classes $i$ and $j$. \\ 
 		\\
 		$\sigma_i$    	    &  rate at which exposed of class $i$ convert into the infected class.\\ 
 		\\ 
 		$\gamma_i $     &  class $i$ host's recovery rate. \\ 
 		\\
 		$ m_i$      	&  host's pathogen-induced death  rate at class $i$. \\ 
 		\\ \hline
 	\end{tabular}%
\end{table}
If it is assumed that the demographic, disease and quarantine parameters are equal for all age classes, the above system reduces to the classical SEIR system with the quarantine term as suggested by \cite{jia2020modeling}. This will be important for what follows.
The parameters for the classical SEIR model will be estimated so that the number of cases predicted by the model  compares well
with the data. This set of parameters will be used later to adjust the age-structured model \ref{eq:seir_estruturado}.

\section{The unstructured SEIR model}
\label{section:The unstructured SEIR model}
\begin{figure}[!h] 
	\centering
	\includegraphics[scale=0.5,trim={2.0cm 7.0cm 0.5cm 7.0cm}]{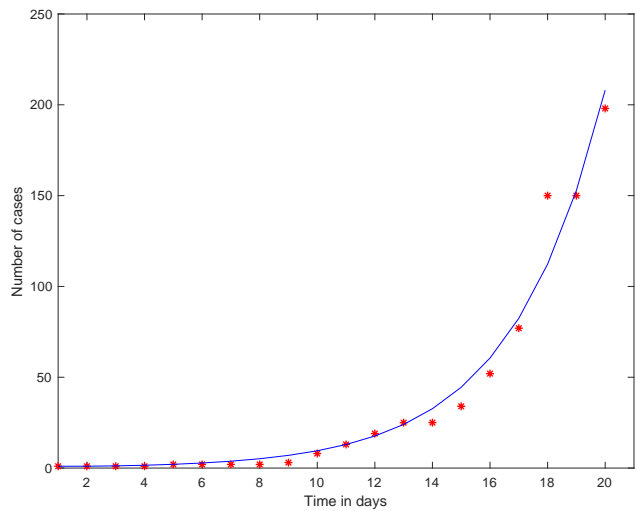}
	\caption{The number of infected for the SEIR non-structured model. The used parameters values are as in \ref{eq:ajustados} . }
	\label{fig:Ajuste}  
\end{figure}

The SEIR model without vital dynamics and with quarantine terms is given by
\begin{equation}
\begin{aligned}
\label{eq:sirbasico_estruturado}
Q^{\, \prime}(t)&= p \, S(t) - \lambda \, Q(t) \, ,  
\\[0.4cm]
S^{\, \prime}(t) &=  -   \beta \, S(t) \, I(t) - p \, S(t) + \lambda \, Q(t) \, ,\\[0.4cm]
E^{\, \prime}(t) &=   \beta \, S(t) \,  \, I(t) - \sigma\, E(t) \, ,\\[0.4cm]
I^{\, \prime}(t) &=  \sigma \, E(t)  - \gamma  \, I(t) \, \, ,\\[0.4cm]
R^{\, \prime}(t) &= \gamma \, I(t)  \, . \\
\end{aligned}
\end{equation}
Ignoring the quarantine class ($p=\lambda=0$) the parameters $\beta$, $\sigma$ and $\gamma$
can be adjusted so that the SEIR curve fits the data. To achieve that the difference between the SEIR infected curve and the data curve for the number of infected is minimized (see \cite{martcheva2015introduction} for algorithm description). The parameters found were   
\begin{equation}
 \label{eq:ajustados}
 \beta^* = 0.8481, \, \ \,  \, \, \, \sigma^* = 0.2682 \, \, \, \mbox{and} \, \, \,  \gamma^* = 0.0870 \, . \end{equation}
 The used initial conditions for the algorithm were
 $ \beta = 2.2 / 2.9$ , $\sigma = 1/ 5.2$ and $\gamma = 1 / 2.9. $
 The  figure \ref{fig:Ajuste} shows the data and the SEIR infected curve using the parameters from \eqref{eq:ajustados}. The considered time interval was 20 days.
\begin{Remark}
The model must be considered with care. The curve $I(t)$, as given by the SEIR model, predicts the total number of infected individuals (symptomatic and asymptomatic) at time $t$.  However, to estimate the number of individuals that will need medical care, one needs to know the proportion between the reported and unreported cases. Estimates for the number of unreported cases can be found at  \cite{russel} and the severity of the reported cases can be found at  \cite{esp}.
Asymptomatic cases can be as high as $75 \%$ \cite{day} of all cases; also, ratio estimates of reported to non reported cases goes from $1/1$ to  $1/20$ \cite{russel}. 
These  uncertainties  must be taken into consideration when  using the model to make numerical previsions. The emphasis of this paper is placed on  understanding qualitatively efficient ways of controlling the epidemic.   \end{Remark} \par 
  Quarantines will be characterized by two values: the entrance rate $p$ and  the exit rate $\lambda$. $p$ is composed of two terms, $\gamma_q$ and $\xi$. $\gamma_q$ is the average time it takes for a person to enter quarantine (see \cite{jia2020modeling}) and $\xi$ is a dimensionless multiplicative factor representing the percentage of individuals that in fact voluntarily quarantine.  With this notation 
 $$p= \frac{\xi}{\gamma_q}\, . $$
  As an example, suppose that $70 \%$ of the population quarantine in an interval of 2 days. Then $p= 0.70 / 2 = 0.35 $. It will be assumed that $  p \in [0.0 \,  , \, 0.40 ] \, .$ $p=0$ means that there is no quarantine.  As in \cite{jia2020modeling} it will be assumed that the time to leave quarantine will between 30 and 60 days; giving that $\lambda \in [1/60 \, , \, 1/30] $. \par
\begin{figure}
\centering
\begin{minipage}{.5\textwidth}
  \centering
  \includegraphics[width=.4\linewidth,trim={8.0cm 8.0cm 8.0cm 8.0cm}]{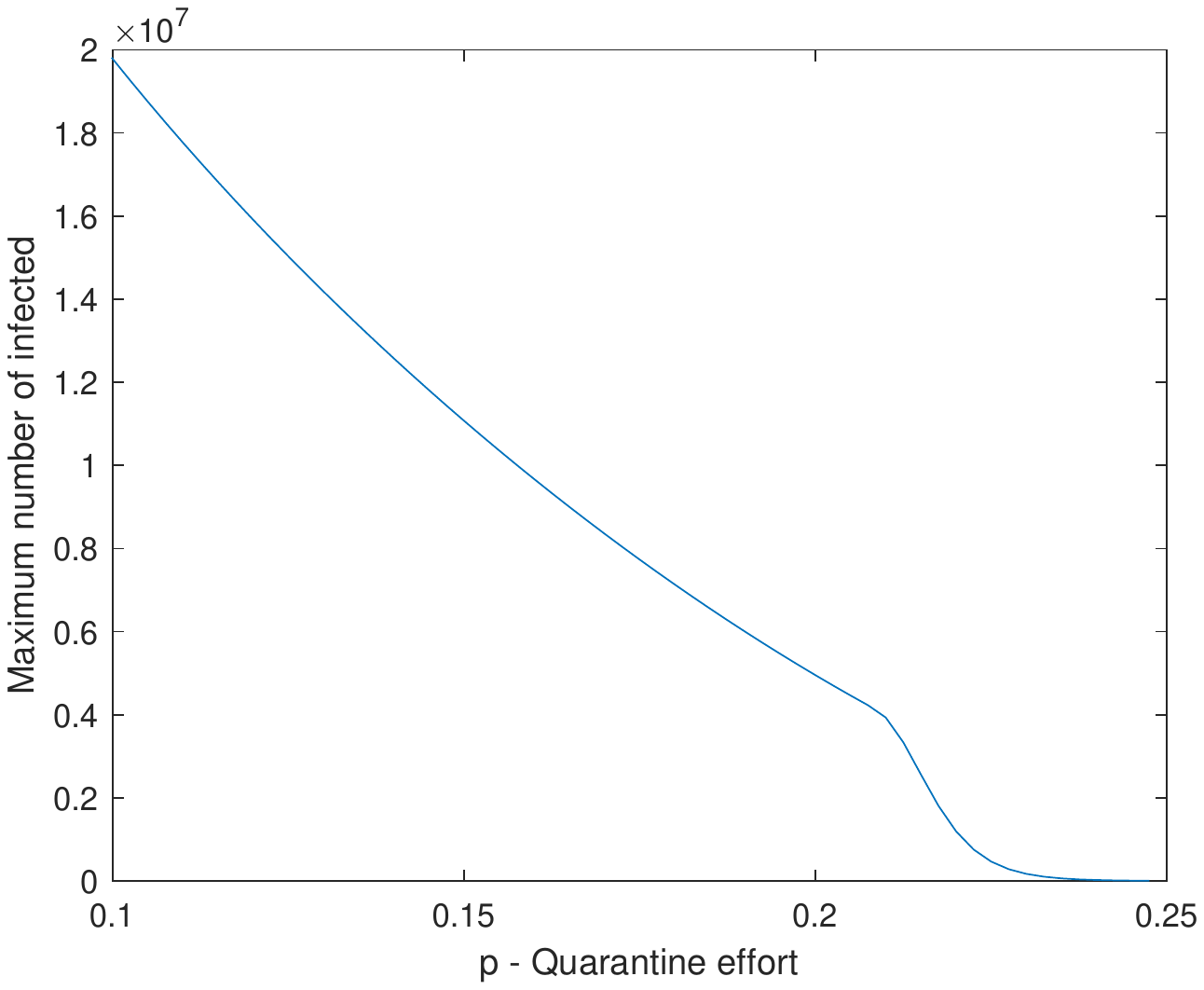}
\end{minipage}%
\begin{minipage}{.5\textwidth}
  \centering
  \includegraphics[width=.4\linewidth,,trim={8.0cm 8.0cm 8.0cm 8.0cm}]{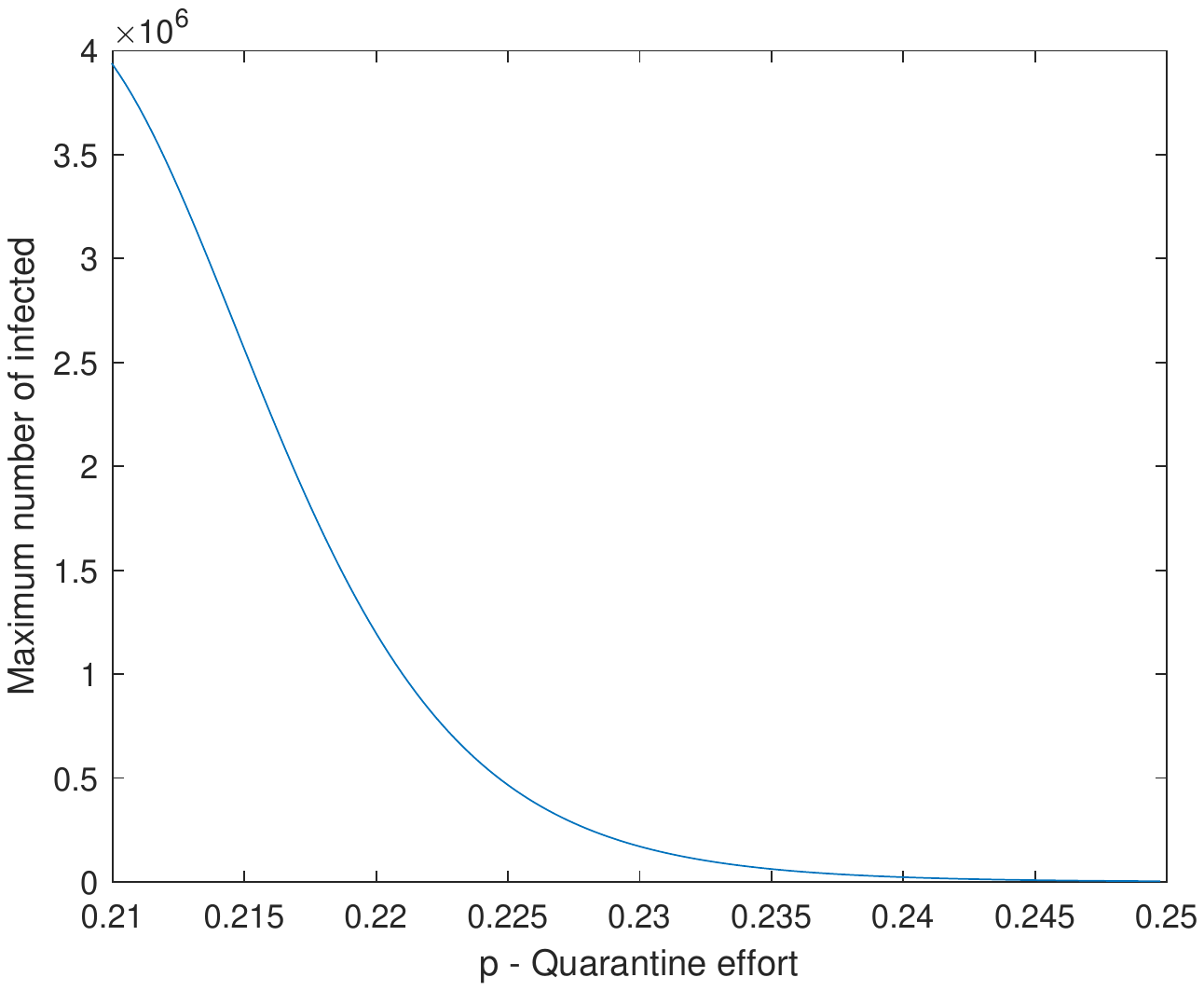}
\end{minipage}
\caption{Maximum of infected as a function of the quarantine effort $p$. The figure on the right-hand side details the fast decrease after $p=0.21$ . Quarantine time is $30$ days. }
\label{fig:SEIR_maximo_por_esforco}
\end{figure}

 
 \begin{Remark}
 \label{rmk:p_menor}
For future reference we observe that, from definition, $p$ is smaller then the percentage of quarantined population.
 \end{Remark}
\begin{figure}
\centering
\begin{subfigure}[b]{0.55\textwidth}
   \includegraphics[scale=0.6,trim={5.0cm 8.0cm 7.0cm 7.0cm}]{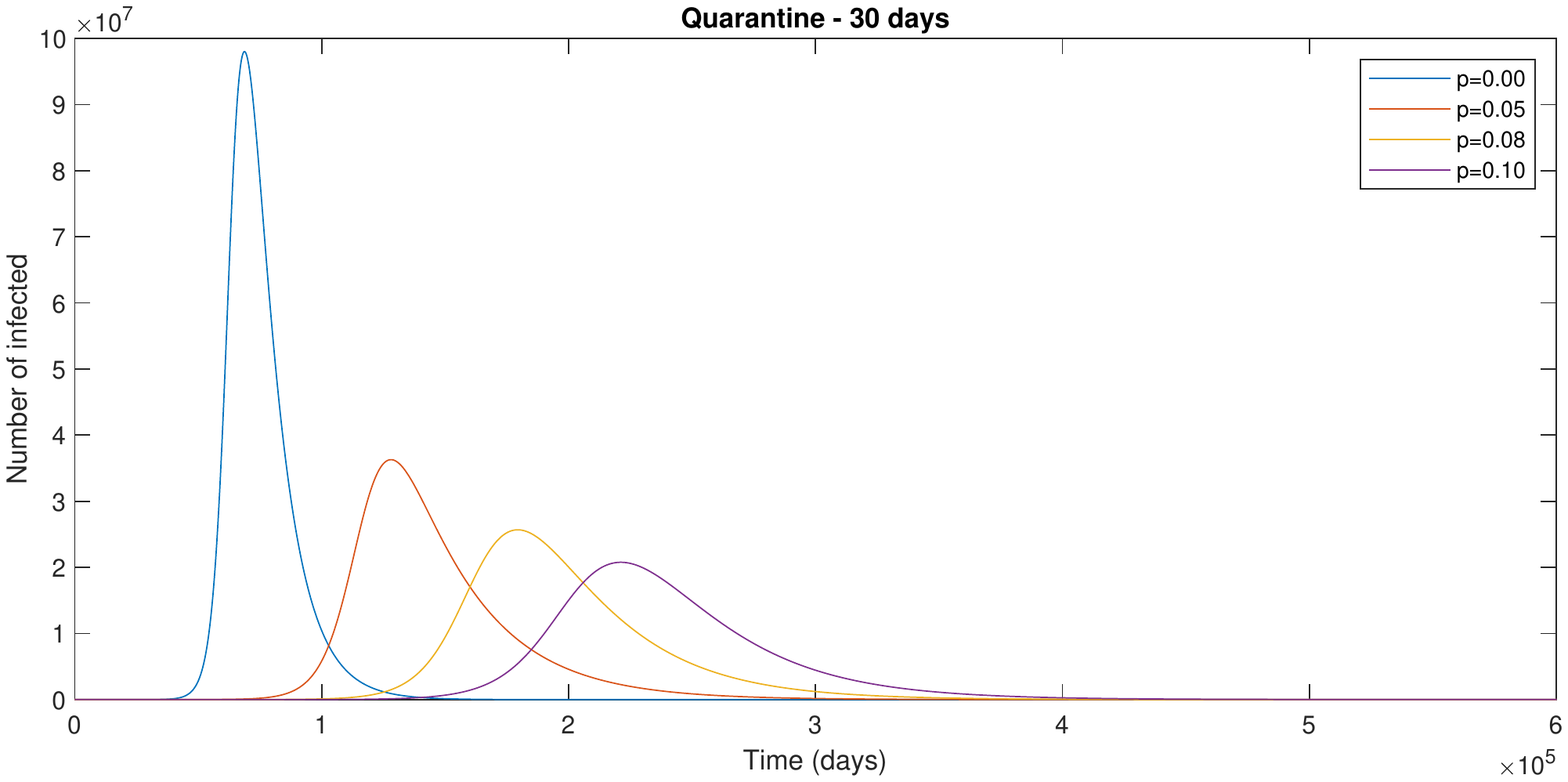}
   \label{fig:SEIR_simples_esforco_1} 
\end{subfigure}
\begin{subfigure}[b]{0.55\textwidth}
   \includegraphics[scale=0.6,trim={5.0cm 10.0cm 0.0cm 7.0cm}]{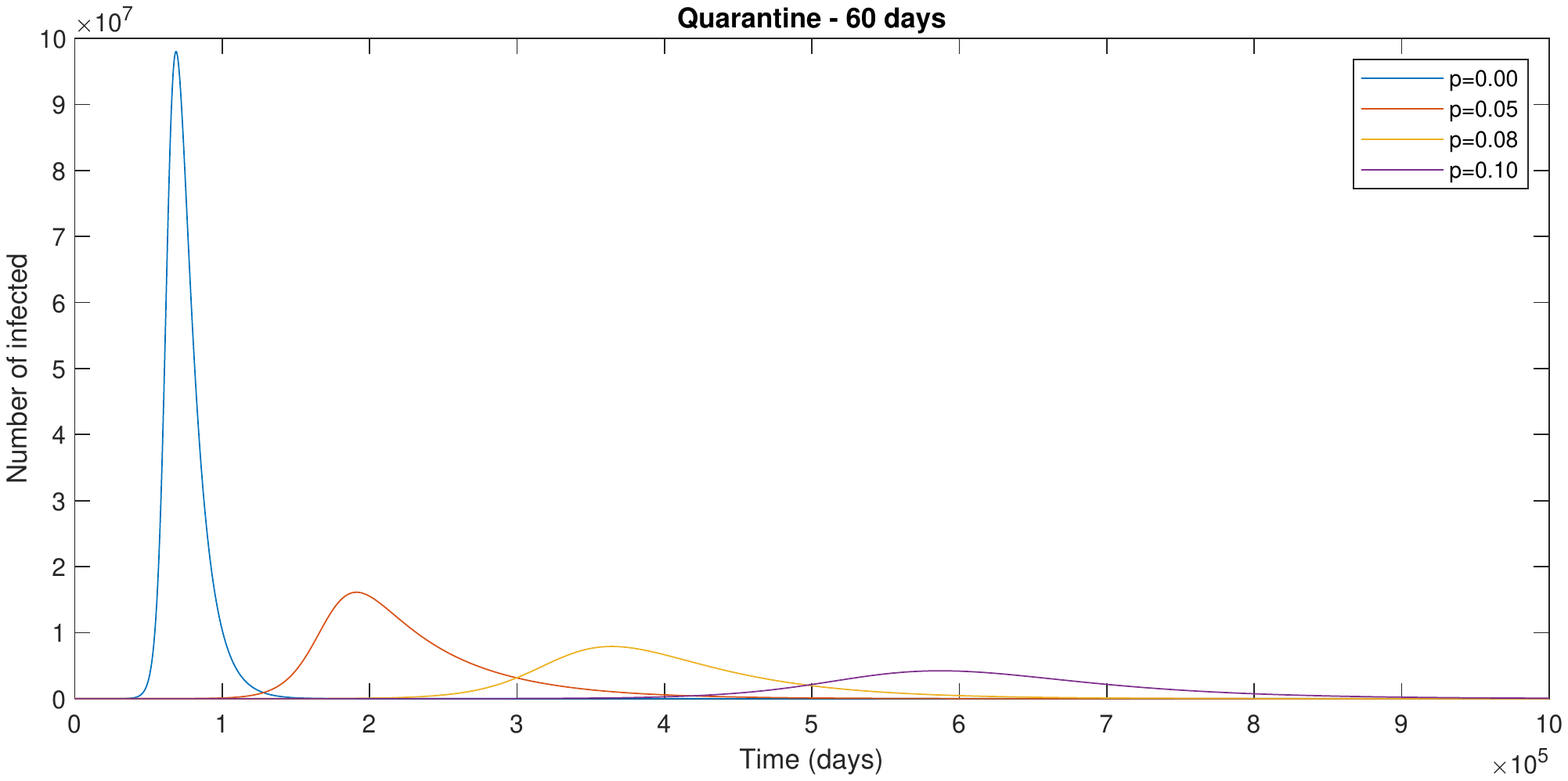}
   \label{fig:SEIR_simples_esforco_2}
\end{subfigure}
\caption{Prevalence curve for different quarantine efforts. The top figure assumes a 30 days quarantine and the bottom figure a 45 days quarantine. }
 \end{figure} 
   The effect of the quarantine  on the prevalence curve is twofold: it decreases the maximum  $I(t)$ value and postpones the date of its occurrence. To assess the efficiency of the quarantine, the maximum of the prevalence curve and the time of its occurrence were calculated and are shown in Figure \ref{fig:SEIR_maximo_por_esforco} .

\begin{figure}[!h] 
	\centering
	\label{fccs}     
	\includegraphics[scale=0.6,trim={2.0cm 8.0cm 0.5cm 8.0cm}]{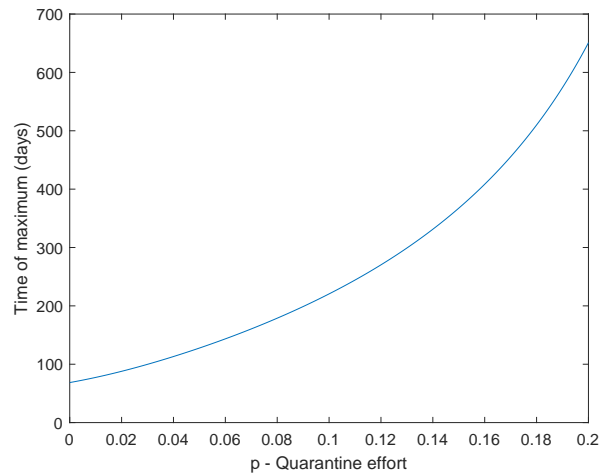}
	\caption{The time for the maximum of the epidemic curve as a function of the quarantine effort $p$. Quarantine time is 30 days.}
\end{figure}
\begin{figure}[!h] 
	\centering
	\includegraphics[scale=0.6,trim={0.0cm 8.0cm 0.0cm 7.0cm}]{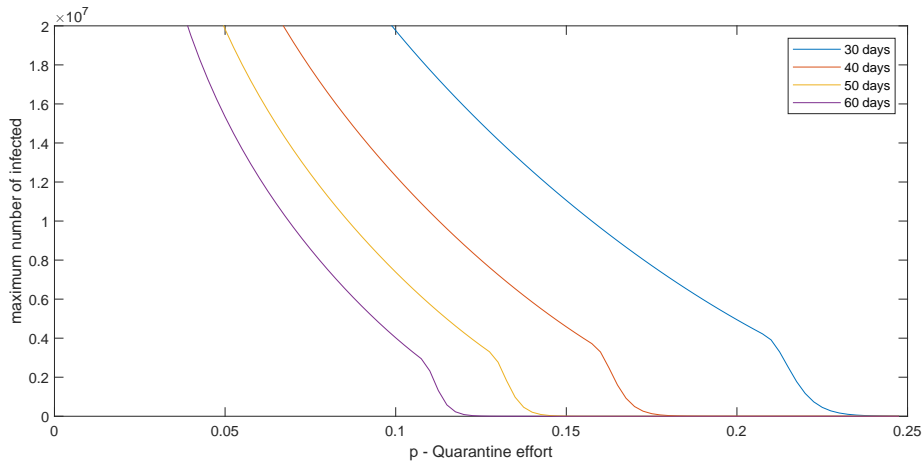}
	\caption{The maximum of the epidemic curve as a function of the quarantine effort $p$ for different quarantine values.}
	\label{fig:SEIR_simples_threshold}  
\end{figure}

 
The important feature on Figure \ref{fig:SEIR_maximo_por_esforco} is the existence of a threshold value for the epidemic effort. For values greater than this critical value, the maximum number of infected decreases extremely fast and the maximum time essentially stabilizes. This is a common feature for all $p$ and $\lambda$ as shown in figure \ref{fig:SEIR_simples_threshold}. 

\begin{figure}[!h] 
	\centering
	\includegraphics[scale=0.6,trim={0.0cm 8.0cm 0.0cm 7.0cm}]{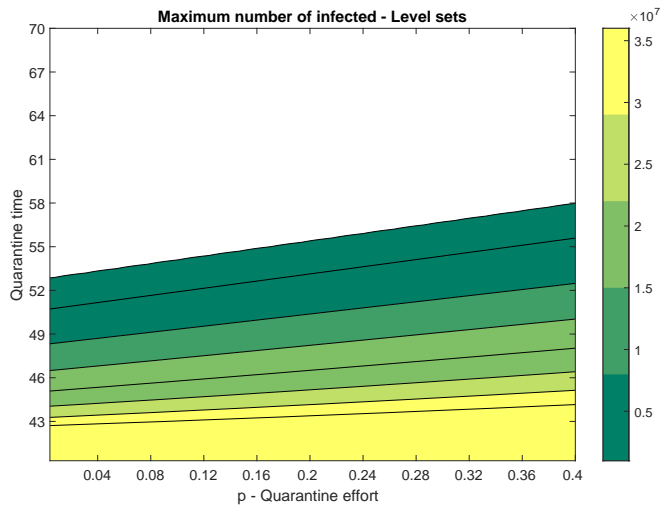}
	\caption{Level curves for the maximum number of infected as a function of quarantine effort and quarantine time.}
	\label{fig:SEIR_simples_esforco_maximo}  
\end{figure}

Critical values for quarantine efforts are clearly seen for the contour plots for the maximum number of infected. The white region on Figure \ref{fig:SEIR_simples_esforco_maximo} divides the parameter plane in two regions. The region above has a maximum number of infected smaller then $ 1 \times 10^6$ infected (by the above rough estimates $\approx$ 5000 deaths ). The region below has larger numbers of infected (and of deaths). The level sets accumulate around a critical level set, showing again that, qualitatively, quarantines do work.

\newpage 
  \section{Control strategies for the age-structured model}
  \label{section:Control strategies for the age-structured model}
  The control measures for the age structured model will be divided in two 
  types. The first  type  controls the epidemic parameters (\cite{castillo1989epidemiological}). This will be made through an $R_0$ sensitivity analysis: the $R_0$ for the age structured model will be numerically determined  and its parameter dependence will be investigated.  The second type of control will be the age-oriented quarantines. The parameters $p_i$ determine the quarantine effort for each class. Due to the different classes weight on the population composition, and to the different epidemic parameters of each class, this study allows us to assess the impact of each class quarantine on the epidemic dynamics. \par 
  Before we proceed, we need to adjust the parameters for the structured model.
  
  \subsection{Data Fitting}
  There are 12 parameters to be determined for the structured model:
  $$ \sigma_i \, \, \, , \, \, \,  \gamma_i \,  \, \, \mbox{and} \, \,  \, 
  \beta_{ij} = \beta_{ji} \, \, \, \mbox{for} \, \, \, i,j = 1,2,3 \, .$$
  The algorithm fits the parameters to the available data of Brazil's total number of reported cases for the first 19 days of infection by a least squares method. The distance between the predicted curve
  $$ I(t) = I_1(t) + I_2(t) + I_3(t) $$
  and the data curve is minimized. The initial parameters for the minimization search algorithm  are based on the ones found for the unstructured SEIR model (\ref{parameters-struc}) taking into consideration the population percentage of each age class. Let $c_i$ be the population percentage of each class, that is (see Table \ref{t-age_classes}), $c_1=0.402$, $c_2=0.505$ and $c_3=0.093$. The initial values for the interation are chosen as
  $$ \sigma_i = c_i \, \frac{\sigma^*_i}{3} \, \, \, \mbox{and} \, \, \, \gamma_i = c_i \, \frac{\gamma^*_i}{3} \, \, \, \mbox{for} \, \, \, i=1,2,3$$ 
  and 
  $$ \beta_{ij} = \beta^* \, \, \, \mbox{for} \, \, \, i\le j=1,2,3 \, ,  $$

\begin{figure} 
	\centering
	\includegraphics[scale=0.5,trim={2.0cm 7.0cm 3.5cm 8.0cm}]{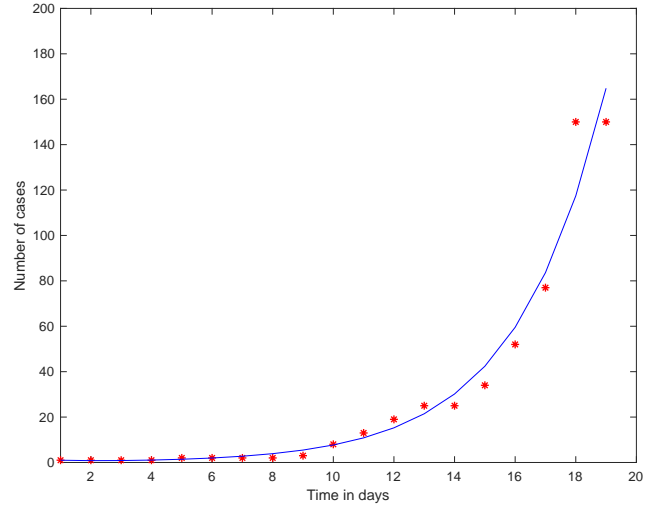}
	\caption{The number infected for the SEIR structured model. Parameter values as in Table \ref{parameters-struc}}
	\label{fig:AjusteParamEst} 
\end{figure}

  The resulting values are listed in Table \ref{parameters-struc} and a plot of the daily number of infections and the number of reported cases is shown in Figure \ref{fig:AjusteParamEst}. 

\begin{table}[!h]
 	\centering
 	\caption{Fitted parameters for the age-structured SEIR model without vital dynamics.}
 	\label{parameters-struc}
 	\begin{tabular}{cccc}
 		\hline  
 		\\
 		Parameter 	  &  Value   & Parameter   & Value\\ 
 		\\ \hline
 		\\

 		$\beta_{11}$ 	    &  1.76168 & $\sigma_1$    	    &  0.27300 \\ 
 		\\
 		$\beta_{12}$    	    &  0.36475 & $\sigma_2$     &  0.58232 \\ 
 		\\ 
 		$\beta_{13}$     &  1.32468 & $\sigma_3$      	&  0.69339\\ 
 		\\
 		$\beta_{22}$      	&  0.63802 & $\gamma_1$   		 &  0.06862 \\ 
 		\\
 		$\beta_{23}$   		 &  0.35958 & $\gamma_2$   		 &  0.03317 \\ 
 		\\
 		$\beta_{33}$ 	    &  0.57347 & $\gamma_3$   		 &  0.35577\\ 
 		\\\hline
 	\end{tabular}%
\end{table}

\newpage 
 \subsection{$R_0$ Analysis}
\begin{figure}
  \includegraphics[scale=0.5,trim={8.0cm 8.0cm 8.0cm 8.0cm}]{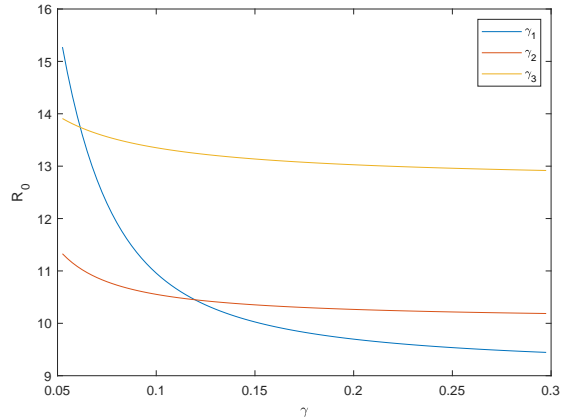}
\caption{$R_0$ as a $\gamma_i$ function. Different curves show which $\gamma_i$ is varying while the others are kept constant.}
\label{fig:SEIR_estruturado_R0_gamma.pdf}
\end{figure}

\begin{figure}
  \includegraphics[scale=0.5,trim={8.0cm 8.0cm 8.0cm 8.0cm}]{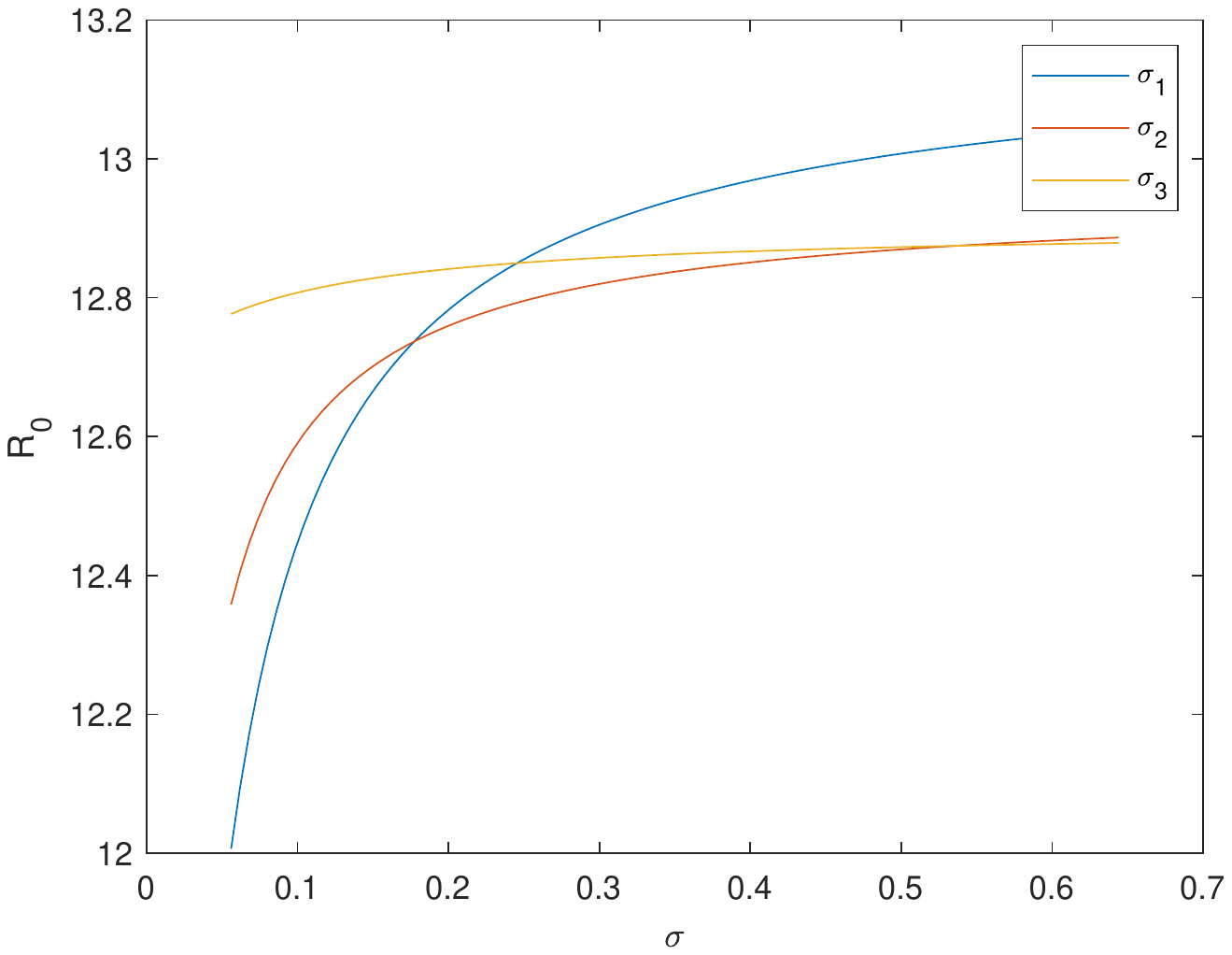}
\caption{$R_0$ as a $\sigma_i$ function. Different curves show which $\sigma_i$ is varying while the others are kept constant.}
\label{fig:SEIR_estruturado_R0_sigma.pdf}
\end{figure}

\begin{figure}
  \includegraphics[scale=0.5,trim={8.0cm 8.0cm 8.0cm 8.0cm}]{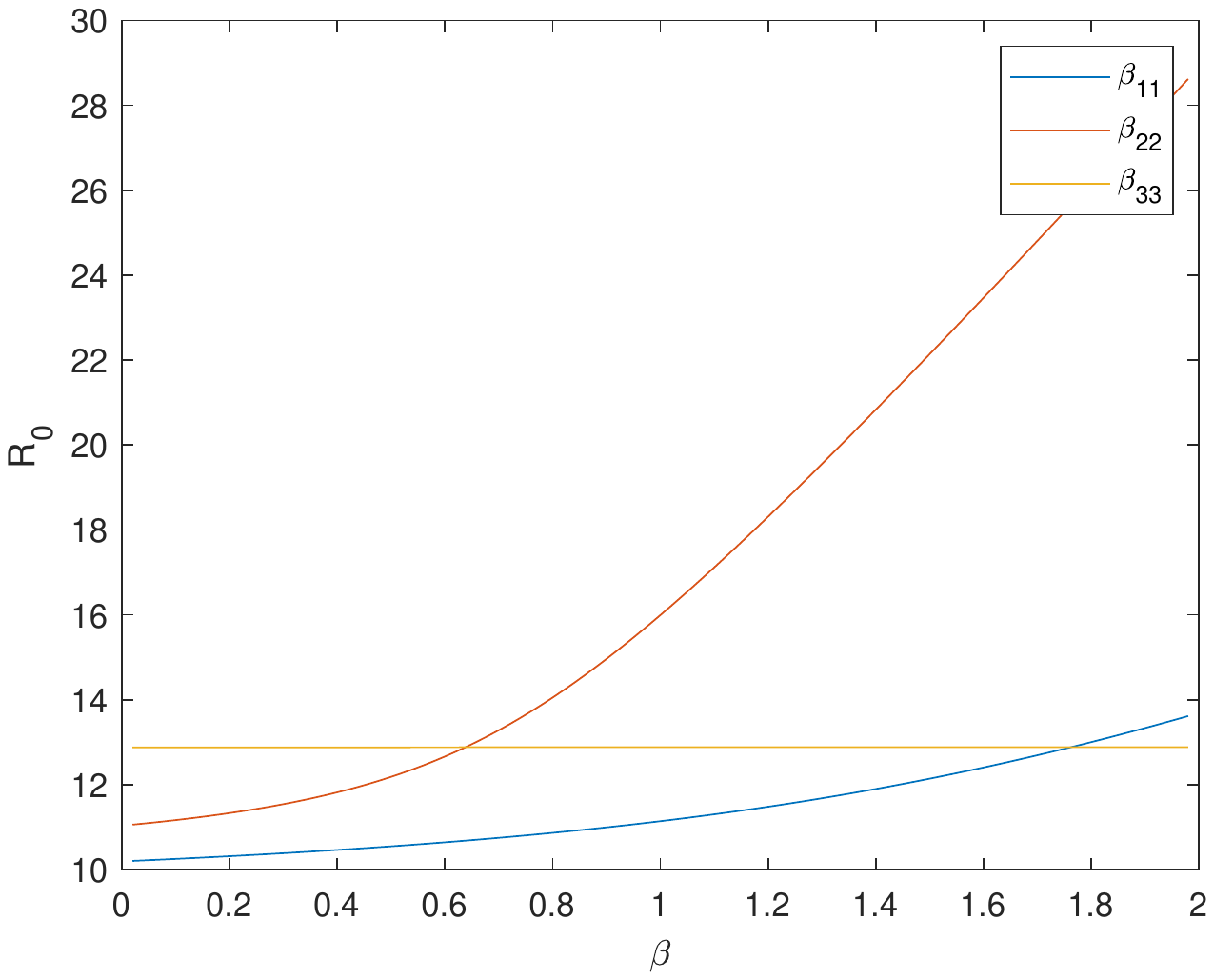}
\caption{$R_0$ as a $\beta_{ii}$ function. Different curves show which $\beta_{ii}$ is varying while the others are kept constant.}
\label{fig:SEIR_estruturado_R0_beta_diagonal.pdf}
\end{figure}

\begin{figure}
  \includegraphics[scale=0.5,trim={8.0cm 8.0cm 8.0cm 8.0cm}]{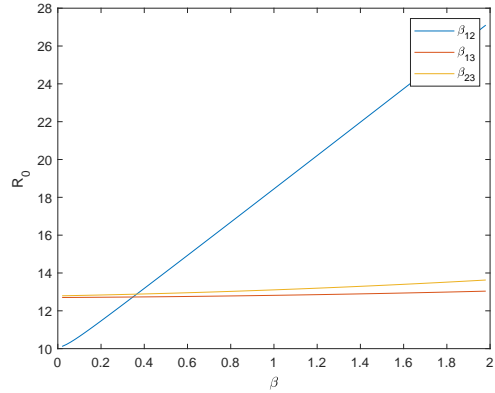}
\caption{$R_0$ as a $\beta_{ij} \, \, i < j \, $ function. Different curves shows which $\beta_{ij}$ is varying while the others are kept constant.}
\label{fig:SEIR_estruturado_R0_beta_off_diagonal.pdf}
\end{figure}

$R_0$ is determined via the next generation approach \cite{diekmann1990definition}. It equals the spectral radius of $FV^{-1}$, where

$$ F= \left( \begin{matrix}
	0 & 0 & 0 & \beta_{11} \, S_1^* &  \beta_{12} \, S_1^* &  \beta_{31} \, S_1^* \\
	\\
	0 & 0 & 0 & \beta_{21} \, S_2^* &  \beta_{22} \, S_2^* &  \beta_{23} \, S_2^* \\
	\\
	0 & 0 & 0 & \beta_{31} \, S_3^* &  \beta_{32} \, S_3^* &  \beta_{33} \, S_3^* \\
	\\
	0 & 0 & 0 & 0 & 0 & 0 \\
	\\
	0 & 0 & 0 & 0 & 0 & 0 \\
	\\
	0 & 0 & 0 & 0 & 0 & 0 \\
	\end{matrix} \right) 
$$
and
$$
V= \left( \begin{matrix}
D_1 & 0 & 0 & 0 & 0 & 0 \\
\\
0 & D_2 & 0 &  0 & 0 & 0 \\
\\
0 & 0 & D_3  &  0 & 0 & 0 \\
\\
-\sigma_1 & 0 & 0 & \tilde{D}_1 & 0 & 0 \\
\\
0 & -\sigma_2 & 0 & 0 & \tilde{D}_2 & 0 \\
\\
0 & 0 & -\sigma_3 & 0 & 0 & \tilde{D}_3 \\
\end{matrix} \right), 
$$
where $D_i = \sigma_i + \mu_i$ and $\tilde{D}_i = \gamma_i + \mu_i + m_i$ for $i \in \{1,2,3\}$. Thus, $F \, V^{-1}$, is given by 

$$FV^{-1} = \left( \begin{array}{cc} K_{11} & K_{12} \\ K_{21} & K_{31} \\ \end{array} \right),$$ where the block $K_{11}$ is

$$
\left( \begin{matrix} \frac{\beta_{11}\, \sigma_1 \, S_1^*}{\gamma_1 \, (\sigma_1 + \mu_1) (\gamma_1 + \mu_1 + m_1)} & \frac{\beta_{12}\, \sigma_2 \, S_1^*}{\gamma_2 \, (\sigma_2 + \mu_2) (\gamma_2 + \mu_2 + m_2)}  & \frac{\beta_{13}\, \sigma_3 \, S_1^*}{\gamma_3 \, (\sigma_3 + \mu_3) (\gamma_3 + \mu_3 + m_3)} \\ 
\\ \frac{\beta_{21}\, \sigma_1 \, S_2^*}{\gamma_1 \, (\sigma_1 + \mu_1) (\gamma_1 + \mu_1 + m_1)} & \frac{\beta_{22}\, \sigma_2 \, S_2^*}{\gamma_2 \, (\sigma_2 + \mu_2) (\gamma_2 + \mu_2 + m_2)}  & \frac{\beta_{23}\, \sigma_3 \, S_2^*}{\gamma_3 \, (\sigma_3 + \mu_3) (\gamma_3 + \mu_3 + m_3)} \\
\\ \frac{\beta_{31}\, \sigma_1 \, S_3^*}{\gamma_1 \, (\sigma_1 + \mu_1) (\gamma_1 + \mu_1 + m_1)} & \frac{\beta_{32}\, \sigma_2 \, S_3^*}{\gamma_2 \, (\sigma_2 + \mu_2) (\gamma_2 + \mu_2 + m_2)}  & \frac{\beta_{33}\, \sigma_3 \, S_3^*}{\gamma_3 \, (\sigma_3 + \mu_3) (\gamma_3 + \mu_3 + m_3)} \\ \end{matrix} \right), \\
$$
the block $K_{12}$ is 

$$
\left( \begin{array}{ccc} \frac{\beta_{11}\, S_1^*}{\gamma_1 + \mu_1 + m_1} & \frac{\beta_{12}\, S_1^*}{\gamma_2 + \mu_2 + m_2} & \frac{\beta_{13}\, S_1^*}{\gamma_3 + \mu_3 + m_3} \\ \\ \frac{\beta_{21}\, S_2^*}{\gamma_1 + \mu_1 + m_1} & \frac{\beta_{22}\, S_2^*}{\gamma_2 + \mu_2 + m_2} & \frac{\beta_{23}\, S_2^*}{\gamma_3 + \mu_3 + m_3}  \\ \\  \frac{\beta_{31}\, S_3^*}{\gamma_1 + \mu_1 + m_1} & \frac{\beta_{32}\, S_3^*}{\gamma_2 + \mu_2 + m_2} & \frac{\beta_{33}\, S_3^*}{\gamma_3 + \mu_3 + m_3} \\  \end{array} \right)
$$
and $K_{21}$ and $K_{22}$ are the $3\times 3$ zero matrix. Due to the block structure of $F \, V^{-1}$, its eigenvalues are easily calculated. However, due to the high number of parameters, their expression is too cumbersome to be of any analytical use. The sensitivity analysis is therefore computed numerically. Figures \ref{fig:SEIR_estruturado_R0_gamma.pdf}, \ref{fig:SEIR_estruturado_R0_sigma.pdf}, \ref{fig:SEIR_estruturado_R0_beta_diagonal.pdf} and \ref{fig:SEIR_estruturado_R0_beta_off_diagonal.pdf} show the $R_0$ parameter dependence.


The use of the parameters of the classical SIR model as control variables  was studied at \cite{castilho2006optimal}. We will follow its interpretation. Measures as keeping social distance, wearing protective masks, washing hands, etc have the effect of reducing the contact rates $\beta_{ij}$. Identifying infected through tests, body temperature checks, etc and putting them into quarantine has the effect of increasing the removal rates $\gamma_i$. $\sigma_i$ is a parameter that can not be controlled.\par The results can be summarized as follows:  \par 
i) Class 1 is the most sensitive to screening measures (see figure
\ref{fig:SEIR_estruturado_R0_gamma.pdf} ). Youngsters should be preferentially  screened. \par
ii) Considering the direct contacts within the same class, class 2 is the more sensitive (see Figure
\ref{fig:SEIR_estruturado_R0_beta_diagonal.pdf} ). Social distance between  adults has the biggest impact on $R_0$.\par 
iii) For the direct contact between different class, $\beta_{12}$ has the greatest impact on $R_0$.

\section{The effects of different quarantine policies} 
\label{section:The effects of different quarantine policies}
In this section we study the impacts of different quarantine strategies.
The disease induced mortality rate was taken into account by considering the number of deaths as a fraction of the recovered class. Death rates for all age groups are estimated using the data from \cite{esp} (Table \ref{deathrates}). As mentioned in Section 3, $I_1$, $I_2$ and $I_3$  include symptomatic and asymptomatic infected individuals as well as unreported cases, so death rates will be multiplied by a factor of $0.25$ (since only 25$\%$ of the infected are symptomatic \cite{day}) and by $1/20$ (due to unreported cases \cite{russel}). This leaves us with a multiplicative factor of $\phi = 0.25*(1/20) = 0.0125$ to estimate the number of deaths. Since we will be working with relative proportions, the actual value of $\phi$ will be of no importance.

\begin{table}[!h]
 	\centering
 	\caption{Death rates for the age-structured model (data taken from \cite{esp}).}
 	\label{deathrates}
 	\begin{tabular}{cccc}
 		\hline  
 		\\
 		Age group 	  &  Number of cases  & Deaths  & $\%$ Death \\ 
 		\\ \hline
 		\\

 		1 	    &  350   & 1      &  $0.29 \%$ \\ 
 		\\
 		2       &   9541 &   36   &  $0.38 \%$ \\ 
 		\\ 
 		3       &  9068 & 	768  &  $8.47 \%$ \\ 
 		\\\hline
 	\end{tabular}%
\end{table}

With these values at hand, we can study the impact of a quarantine with parameters $\lambda$ and $p_i$, for $i \in \{1,2,3\}$. Calling $p$ the quarantine effort for the unstructured model, it is assumed that the total quarantine effort  equals the effort for the non-structured model
that is
$$ p_1 + p_2 + p_3 = p\, .$$
\begin{Remark}
\label{rmk:pmenor_classe} As mentioned in \ref{rmk:p_menor}, calling $q_i \, \, \, i=1,2,3$ the percentage of quarantined on each age class, it follows that
$$ p < q_1 + q_2 + q_3 \le 1 \, . $$
\end{Remark}

Four different choices for the $p_i$'s will be used, as detailed in Table \ref{strategies}. These are choices for the quarantine effort of each age group. Strategy S1 splits the effort equally among the three groups. Strategy S2 emphasizes a stronger isolation of the elderly (twice as much as the other groups). Strategy S3  enforces isolation of the youngsters and adults twice as much as it does for the elderly. Finally, strategy S4 doubles the quarantine effort on the adults in comparison to the others.  To assess the efficiency of these different control strategies, for a fixed control effort $p$, each control strategy will be calculated for different quarantine times   $\lambda \in \{1/30, 1/45, 1/60\}$. 

\begin{table}[!h]
 	\centering
 	\caption{Quarantine strategies.}
 	\label{strategies}
 	\begin{tabular}{cc}
 		\hline  
 		\\
 		Strategy	  &  Choices for the $p_i$   \\ 
 		\\ \hline
 		\\

 		S1 	    &  $p_1 = p/3$, $p_2 = p/3$, $p_3 = p/3$ \\ 
 		\\
 		S2      &  $p_1 = p/6$, $p_2 = p/6$, $p_3 = 2p/3$ \\ 
 		\\ 
 		S3      &  $p_1 = 2p/5$, $p_2 = 2p/5$, $p_3 = p/5$ \\
 		\\
 		S4      &  $p_1 = p/6$, $p_2 = 2p/3$, $p_3 = p/6$ \\ 
 		\\\hline
 	\end{tabular}%
\end{table}

The estimation of the number of deaths can be made by multiplying the number of recovered at the end of the epidemic in each of the three age groups by the death rates from Table \ref{deathrates} and by the multiplicative factor $\phi$. However, due to parameters uncertainties and lack of estimations for the parameter $p$, a different approach is taken. We arbitrarily chose one of the values as unit and calculated all the other results proportionally.  The results for $p = 0.2$ are available in Table \ref{results}. (For reference only, the number of deaths chosen as unit was 2869).

\begin{table}[!h]
 	\centering
 	\caption{Proportion of deaths for each age group for different quarantine strategies and durations.}
 	\label{results}
 	\begin{tabular}{cccccc}
 		\hline  
 		\\
 		$\lambda$ 	  &  Age group   & S1    & S2    & S3    & S4   \\ 
 		\\ \hline
 		\\

 			                &  1     & 1     & 1.02  & 0.99  & 1.03 \\ 
 		\\
 		1/30                &  2     & 1.61  & 1.67  & 1.59  & 1.47 \\ 
 		\\ 
 		                  &  3     & 7.20  & 6.43  & 7.46  & 7.51 \\ 
 		\\
 		                  & Total  & 9.81  & 9.12  & 10.04 & 10.01 \\
 		\\\hline
 		\\
 		                  & 1      & 0.95  & 0.99  & 0.93  & 1.01 \\
 		\\
 		1/45                & 2      & 1.51  & 1.60  & 1.47  & 1.29 \\
 		\\
 		                  & 3      & 6.77  & 5.75  & 7.18  & 7.26 \\
 		\\
 		                  & Total  & 9.23  & 8.34  & 9.58  & 9.56 \\
 		\\\hline 
 		\\
 		                  & 1      & 0.90  & 0.96  & 0.88  & 0.98 \\
 		\\
 		1/60                & 2      & 1.41  & 1.54  & 1.36  & 1.14 \\
 		\\
 		                  & 3      & 6.38  & 5.21  & 6.90  & 7.01 \\
 		\\
 		                  & Total  & 8.69  & 7.71  & 9.14  & 9.13 \\
 		\\\hline 
 	\end{tabular}%
\end{table}

\begin{figure}[!h] 
	\centering
	\includegraphics[scale=0.42,trim={2.0cm 5.0cm 3.5cm 5.0cm}]{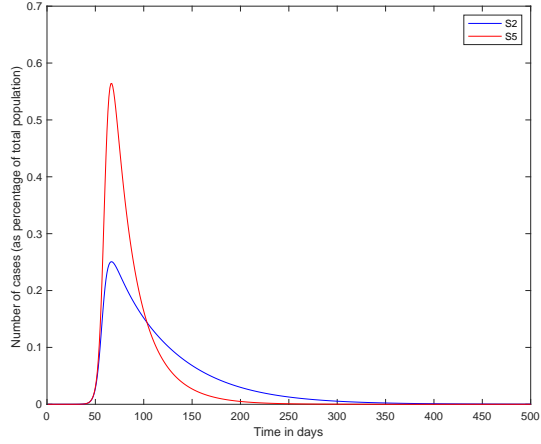}
	\caption{Plots of the total number of infected, as percentages of the total population, for strategies S2 and S5. The quarantine parameters were $p = 0.2$ and $\lambda = 1/60$.}
	\label{fig:infec-estrat} 
\end{figure}

 One could argue that the optimal control would occur if we put all the quarantine effort in the isolation of the elderly and no isolation at all for youngsters and adults. With our terminology, this means considering a strategy S5 defined by $p_1 = p_2 = 0$ and $p_3 = p$. However, this leads to two main problems: first, due to the small percentage of the elder class the quarantine effort would be to small (in fact smaller then 0.1) to be of any significance. Second, it would allow for a much higher number of infected individuals (see Figure \ref{fig:infec-estrat}), hence a great increase in the total of hospitalizations, which would collapse the Health System. Therefore, to achieve better quarantine results, the total effort needs to include all age-groups, with more emphasis on the elderly since they have a higher fatality rate due to the disease. 

\begin{figure}[!h] 
	\centering
	\includegraphics[scale=0.68,trim={2.0cm 6.0cm 3.5cm 4.0cm}]{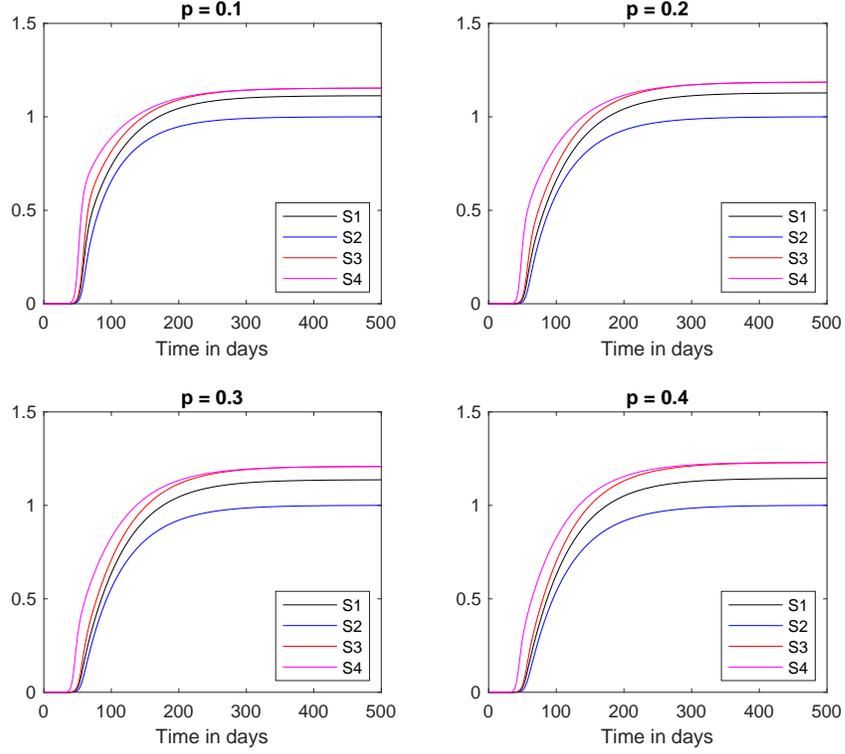}
	\caption{Plots of $\mathcal{D}_j(t)$ for $\lambda = \frac{1}{30}$, $j \in \{1,2,3,4\}$.}
	\label{graph-strat} 
\end{figure}

Notice that strategy S2 is, by far, the best among these. All other strategies end up with, at least, 7.5$\%$ more deaths. We can also analyse the strategies by plotting them. Let $\mu_i$, $i \in \{1,2,3\}$, be the death rates from Table \ref{deathrates}, so $$\mathcal{D}_j(t) = \phi \sum_{i=1}^3  \mu_i R_i(t)$$ converges to the total amount of deaths that result from strategy S$_j$, $j \in \{1,2,3,4\}$. Figure \ref{graph-strat} plots the graphs of $\mathcal{D}_j(t)$, normalized by $$\lim_{t \rightarrow \infty} \mathcal{D}_2(t),$$ produced by the four strategies for different values of $p$. Notice yet that, in all four cases, the strategy that produces the smallest limit value (hence the smallest death toll) is S2.

\section{Conclusions}
\label{ref:conclusions}
In this paper we introduced an age-structured SEIR model with a quarantine compartment.
Three age classes were used: infants (0 to 19 years), adults (20 to 59 years) and elderly (60 to 100 years). First we studied the associated classical unstructured SEIR model without vital dynamics. The parameters were fitted by a least-square algorithm and the impact of the quarantine parameters $p$ and $\lambda$ were studied. Our main findings concern the existence of a numerical threshold value for the quarantine parameters: above a certain curve on the $(p,\lambda)$-plane, the maximum number of infected decreases in an accentuated way. This shows that  an abrupt decline on the number of cases should be observed if the quarantine is being efficient. If this decline is not being observed, quarantine effort and time should be increased. \par 
The parameters obtained for the unstructured SEIR model were used to adjust the parameters for the age-structured SEIR model. Using this data, the basic reproduction number $R_0$ was calculated and its dependence on the epidemic values was studied. Our findings for the $R_0$ analysis are as follows:\par
i) Class 1 is the most sensitive to screening measures (see Figure
\ref{fig:SEIR_estruturado_R0_gamma.pdf}). Youngsters should be preferentially  screened. \par
ii) Considering the direct contacts within the same class, class 2 is the more sensitive (see Figure
\ref{fig:SEIR_estruturado_R0_beta_diagonal.pdf} ). Social distance between  adults has the biggest impact on $R_0$.\par 
iii) For the direct contact between different class, $\beta_{12}$ has the greatest impact on $R_0$.  \par 
Finally we studied the impact of age-oriented campaigns considering different strategies and different values of $p$ for the total campaign effort. Recalling that $p$ is bounded by the percentage of quarantined population (see remarks \ref{rmk:p_menor} and \ref{rmk:pmenor_classe}), our findings show that the highest possible quarantine must be made, and then, this effort must concentrate on putting into quarantine the total of elders and assuring equal proportions of adults and youngsters.
\section*{Acknowledgement} The authors would like to thank Renato Mello (IFSP - Campus Salto) for discussions during the preparation of this manuscript.
\bibliographystyle{amsplain}
\bibliography{references}

\end{document}